\documentclass[preprint,12pt]{emulateapj}

\shorttitle{High Resolution NICE Map of B59}
\shortauthors{Rom\'an-Z\'u\~niga et al.}

\begin{document}

\title{High Resolution Near-Infrared Survey of the Pipe Nebula I: A Deep Infrared Extinction Map of Barnard 59}

\author{Carlos G. Rom\'an-Z\'u\~niga\altaffilmark{1,2,3}, Charles J. Lada\altaffilmark{2} and
Jo\~ao F. Alves\altaffilmark{1,3}}
\altaffiltext{1}{Centro Astron\'omico Hispano Alem\'an, Granada 18006, Spain}
\altaffiltext{2}{Harvard Smithsonian Center for Astrophysics, Cambridge MA 02143}
\altaffiltext{3}{Instituto de Astrof\'isica de Andaluc\'ia, Granada 18006, Spain}

\begin{abstract}
We present our analysis of a fully sampled, high resolution dust extinction map of
the Barnard 59 complex in the Pipe Nebula. The map was constructed with the infrared color excess
technique applied to a photometric catalog that combines data from both ground and space based observations.
The map resolves for the first time the high density
center of the main core in the complex, that is associated with the formation of a small cluster of stars.
We found that the central core in Barnard 59 shows an unexpected lack of significant 
substructure consisting of only two significant fragments. Overall, the material appears to be 
consistent with being a single, large core with a density profile that can be well fit by a King model. A 
series of NH$_3$ pointed observations towards the high column density center of the core appear
to show that the core is still thermally dominated, with sub-sonic non-thermal motions. The stars
in the cluster could be providing feedback to support the core
against collapse, but the relatively narrow radio lines suggest that an additional source of support, 
for example a magnetic field, may be required to stabilize the core.
Outside the central core our observations reveal the structure of peripheral cores and resolve an extended filament into a handful of significant
substructures whose spacing and masses appear to be consistent with Jeans fragmentation.

\end{abstract}

\keywords{infrared: ISM ---  ISM: dust, extinction ---  ISM: globules --- stars:
formation}

\section{Introduction}

Knowledge about the initial conditions of star formation is scarce: in a majority of the observable clouds, star formation has been active for relatively long periods of time, and the clouds have evolved from their early stages. Examples of the few observable clouds that display little or no star formation activity and may be very young, are the Maddalena Cloud \citep{maddalena85}, Lupus 5 \citep{lupus5}, portions of the Aquila Rift \citep{kawamura,lprato} and the Pipe Nebula. The latter, with a mass of $10^4$ M$_\odot$, is particularly interesting because it is one of the nearest clouds \citep[$d=130^{+13}_{-20}$ pc;][]{lal06} that we can observe with a great level of detail. For these reasons the Pipe has recently become a prime target for the study of the early stages of star formation. 

\citet{lal06} (hereafter LAL06) used the observations of the Two Micron All Sky Survey (2MASS) to construct a complete extinction map of the cloud at a resolution of 60$\arcsec$. Using this map, \citet{alola07} and \citet{rath09} (hereafter RLA09) revealed and confirmed, respectively, that the Pipe Nebula hosts over 130 dense cores. According to those studies, the mass spectrum of the Pipe Nebula cores appears to be very similar in shape to the standard initial mass function only scaled to higher mass by a factor of 3. This result appears to suggest that the functional form of the IMF may be imprinted in the primordial fragmentation of the cloud. Radio wavelength observations towards individual cores by \citet{pipec18o} and \citet{pipenh3} allowed them to produce measurements of kinematics, characteristic gas temperatures and gas pressure estimates for a large fraction of the cores. Analysis of combined data indicated that the Pipe Nebula cores are thermally supported, gravitationally stratified and pressure confined entities which are possibly the product of thermal fragmentation \citep{pipecores}. A deep Spitzer Space telescope (SST) mid-infrared (MIPS) survey of the Pipe cloud (Forbrich 2009, in preparation) show that only a handful of the more than 130 dense cores harbor young stellar objects, demonstrating that the vast majority of these cores are starless objects

One core that stands out from the set is Barnard 59 (B59), the largest and most massive core in the cloud. B59 is also one of only a few cores in the nebula with evidence for active star formation, and
the only one forming a group of young stellar objects. Recent SST mid-infrared (IRAC) observations revealed that B59 is forming a small star cluster with 14 potential members \citep[out of at least 20 YSO candidates;][hereafter BHB07]{brooke07}. Most likely the cluster is forming only low mass ($<2\mathrm{\ M}_\odot$) stars. The current star forming efficiency in the cluster would be relatively high, possibly above $\sim 20$\%.

Currently available maps of B59 are insufficient for detailed study of the core structure. The arcminute resolution extinction map of LAL06 could not resolve the central core that hosts the cluster because the 2MASS observations are too shallow to penetrate the highest extinctions in its central region. Currently available radio maps of B59 \citep{onishi99,lohr07}, also lack the resolution to resolve the internal structure of the core. With the goal of making a higher resolution, fully sampled extinction map, we combined Spitzer observations ---that penetrate the very dense center of the core up to a maximum column density of $A_V\sim $100 mag--- with near-infrared (NIR), high resolution observations obtained at 4 and 8m class telescopes of the European Southern Observatory (ESO). Our combined catalog allowed us to determine the near to mid-infrared (MIR; 1 to 8 $\mu$m) extinction law towards B59 to an unprecedented depth of almost 60 visual magnitudes of extinction \citet{b59elaw07}. We found that the resultant extinction law: 1) is approximately linear down to high levels of reddening, i.e., does not change significantly with depth; 2) departs (flattens) only moderately from the extinction law calculated with similar methods towards other less dense regions \citep{indebe05, flaherty07}. These results suggest that  we can reliably measure high dust column densities in B59 using a single extinction law.

\begin{figure}[htbp]
\begin{center}
\includegraphics[angle=0,width=3.5in]{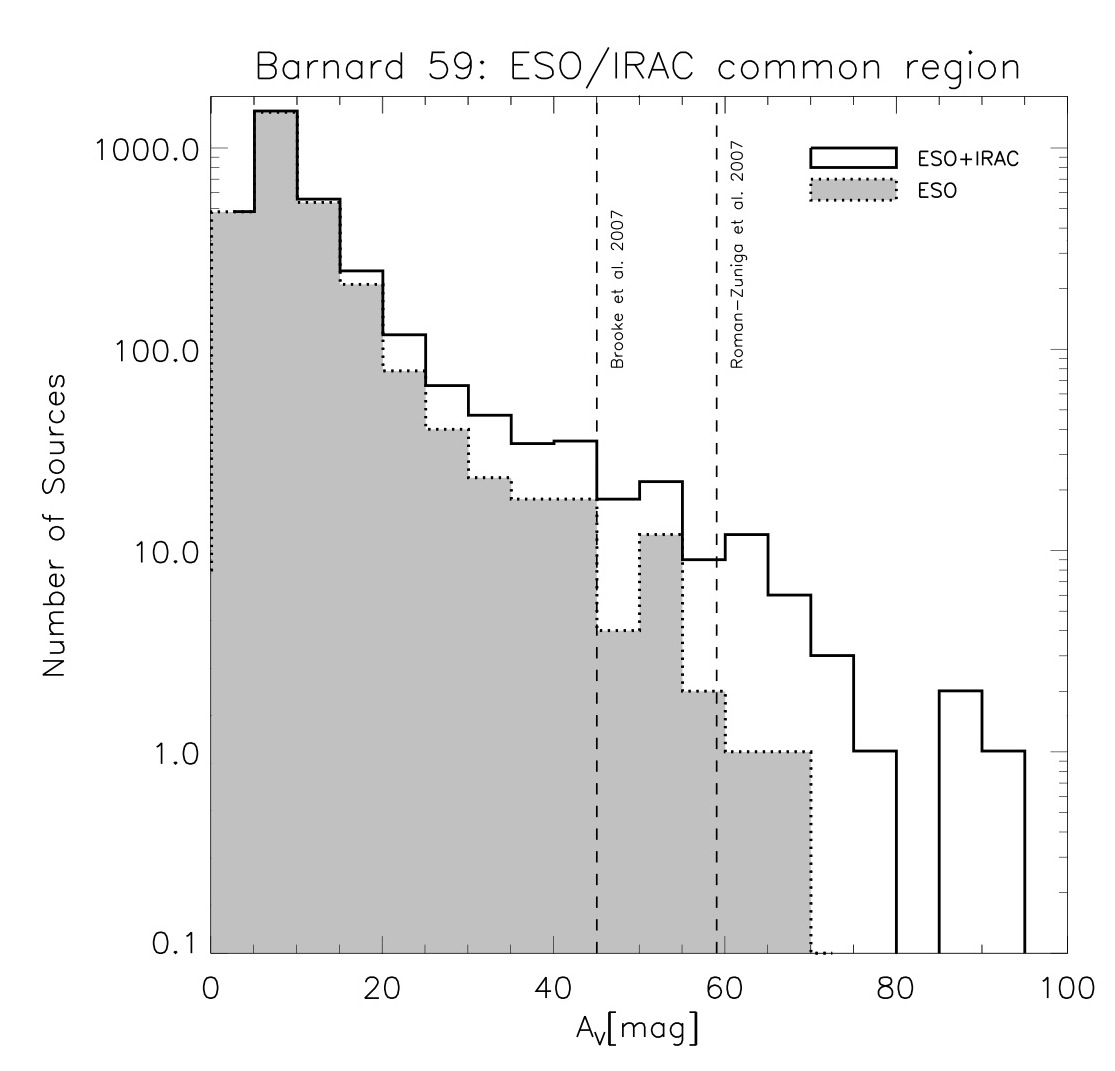}
\caption{Distribution of visual extinction towards Barnard 59. The solid histogram outlined by a dotted line represents extinction values calculated with ESO NIR color excess, while the solid line histogram represents extinction values calculated with color excess from a combined ESO+IRAC catalog. The two vertical lines represent limits used in previous studies: Brooke et al. 2007 (2MASS+IRAC) and Rom\'an-Z\'u\~niga et al. 2007 (ESO+IRAC).\label{fig:exthist}}
\end{center}
\end{figure}

The exercise of \citeauthor{b59elaw07} gave us the confidence to construct a deep extinction map with the combined ESO-SST catalog using the standard color excess method \citep{NICER}. The addition of mid-infrared colors adds the necessary power to resolve the central part of the region, obscured in the NIR. We illustrate this in Figure \ref{fig:exthist}, where we compare measurements of dust extinction towards B59 using near-infrared ($E(H-K)$) color excess alone, and using ``hybrid'' NIR+MIR or ``pure'' MIR color excesses ($E(m_{\lambda _i}-K)$, $E(m_{\lambda _i}-m_{\lambda _j})$; $\lambda=[3.6],[4.5],[5.8],[8.0]\ \mu$m). 

The main goal of this study is to construct a deep extinction map of Barnard 59, that resolves its central core, and use it as a base to further investigate the properties of this isolated cluster forming region. The high depth and resolution of the observations used in this study allow us to investigate the morphology and basic physical characteristics of the central core and its peripheral substructures to an unprecedented level. 

The present paper is organized as follows: in section $\S$\ref{section:observations}, we briefly describe the characteristics of the various sets of data we use. In $\S$\ref{section:analysis} we describe the technique used to construct a deep, infrared dust extinction map of the Barnard 59 complex. In $\S$\ref{section:identification} we describe the analysis performed on the map to identify significant extinction features in it. In section $\S$\ref{section:discussion} we present a discussion on the physical properties of significant features, their association with previously identified cloud cores and the implications for the process of fragmentation in the B59 complex. Finally, in $\S$\ref{section:summary} we make a brief summary of our findings.

\section{Observations}
\label{section:observations}

\begin{figure*}[htbp]
\begin{center}

\includegraphics[angle=0,width=6.0in]{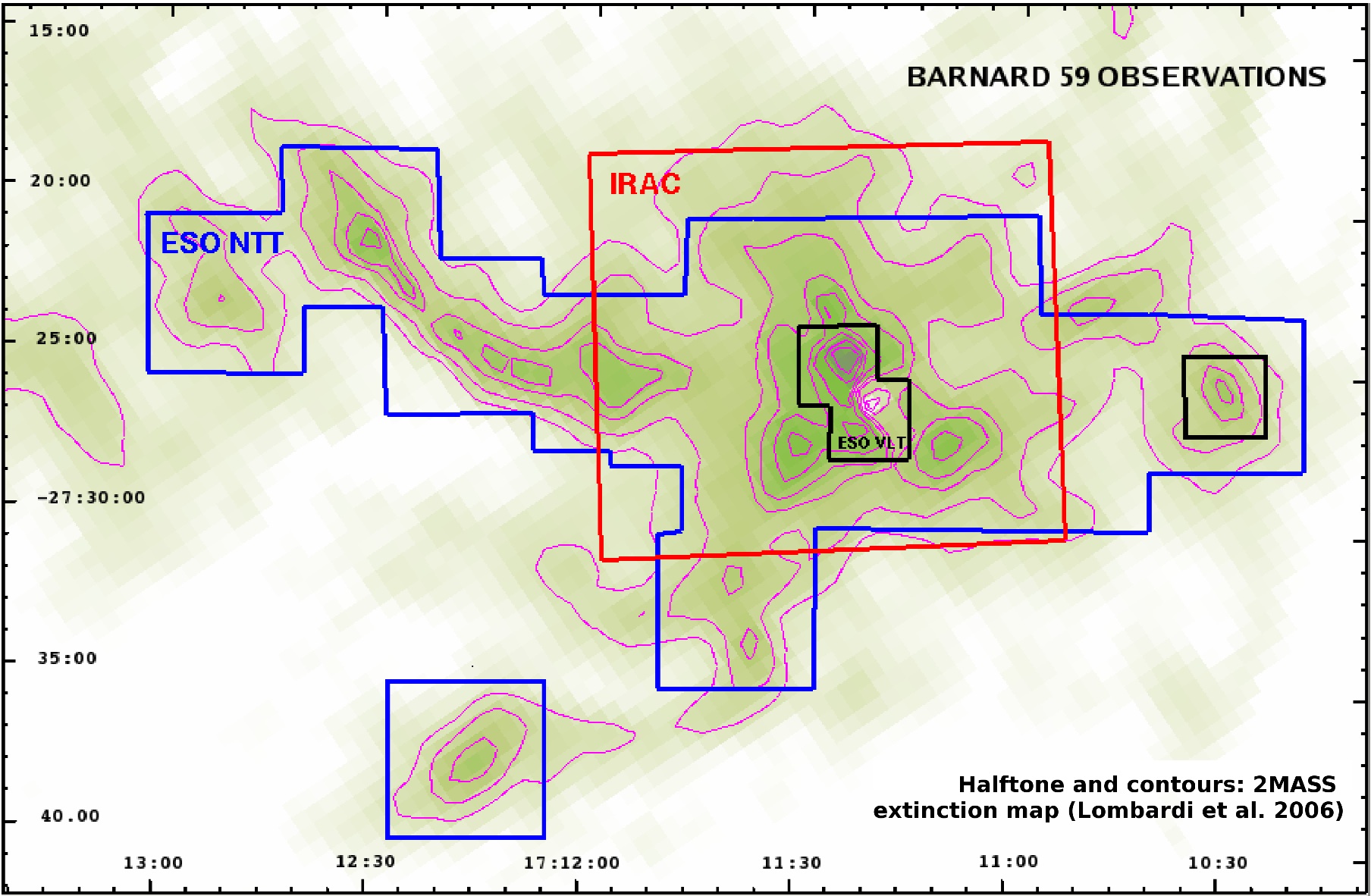}

\caption{Common observations of B59. The area outlined in red denotes the coverage of IRAC. The areas outlined in blue indicates the spawn of the NTT observations. The small areas outlines in black indicate the fields observed with the VLT. The halftone and contours in the background indicate extinction $A_V$ at levels of 5, 7, 9, 12, 15, 18, 20 and 22 mag, as determined by \citet{lal06}. \textit{Please see the electronic version of the Astrophysical Journal for a full color version of this figure} \label{fig:obsmap}}
\end{center}
\end{figure*}

Our investigation was carried out with a combination of ground and space based observations. The coverage of the different datasets is outlined in Figure \ref{fig:obsmap}.

The ground based observations are part of a high resolution NIR survey of regions across the whole Pipe Nebula. The fields corresponding to the Barnard 59 area, which concern this study, are divided in two sets. The first set are NIR ($H$, $K_s$) images obtained with the Son of ISAAC (SOFI) camera at the New Technology Telescope (NTT) facilities of the European Southern Observatory (ESO) atop Cerro La Silla in Atacama, Chile. These SOFI-NTT observations were performed at two different epochs: the first season ran during June 2001, with 8 fields imaged in $H$ and $K_s$; the second season was held in June 2002, with 9 fields imaged in $H$ and $K_s$. In addition, 4 fields of the second season also have coverage in $J$. The survey aimed to cover as much as possible of the main structures visible in early 2MASS extinction maps. The resolution of the SOFI NTT observations is 0.288$\arcsec$. 

The second set of ground-based data are NIR $H$ and $K_s$ images obtained with the Infrared Spectrometer And Array Camera (ISAAC) at the Very Large Telescope (VLT) facilities of ESO atop Cerro Paranal in Antofagasta, Chile. These ISAAC-VLT observations were carried out during July 2002 with integrations of 220 and 390 sec. in $K_s$ and $H$, respectively. These observations targeted the areas of highest extinction in the region, specifically the nucleus of the dense, star forming core (2 fields), and the westernmost core of the cloud (core 03 in the list of RAL09, 1 field). The high resolution ($0.148\arcsec$) and sensitivity of these observations allowed us to detect a few more highly reddened sources that could not be detected with SOFI.

\begin{deluxetable}{lccc}
\tablecolumns{7}
\tablewidth{0pc}
\tablecaption{Color Selection for Hybrid NICE $A_V$ Evaluations \label{tab:colors}} 
\tablehead{
\colhead{Available} &
\colhead{Color} &
\colhead{$C_{el}$} &
\colhead{Intrinsic Color} \\
\colhead{Photometry} &
\colhead{$\lambda _1-\lambda _2$} &
\colhead{} &
\colhead{$(\lambda _1-\lambda _2)_0$} \\
}
\startdata

$H$ and $K_s$ available           & $H-K_s$       &  16.23  &  0.185\\
Null or Poor\tablenotemark{a} $H$ & $[3.6]-K_s$   & -23.41  & -0.528\\
Null or Poor $H$ and $K_s$        & $[3.6]-[4.5]$ &  89.29  &  0.047\\
Null or Poor $K_s$ and $[3.6]$    & $[4.5]-[5.0]$ &  153.96 &  0.129\\
\enddata
\tablenotetext{a}{Photometric uncertainty larger than 0.1 mag} 
\end{deluxetable}

Data reduction is described in detail in a following paper involving the complete ESO NIR survey of the Pipe Nebula (Rom\'an-Z\'u\~niga et al 2009, in preparation). In a nutshell, data reduction was carried out using a pipeline \citep{mythesis} that combines IRAF routines and ESO publicly available routines to take care of particular aspects of the VLT and NTT data processing. A second pipeline \citep{levinethesis} combines the S-Extractor algorithm \citep{sextractor} with IRAF routines to perform Point Spread Function photometry on the crowded fields of the Pipe Nebula, projected against the Galactic bulge.

The SST dataset corresponds to the publicly available pipeline catalog of B59 which is part of the Cores to Disks (C2D) Legacy Project and includes photometry in the four IRAC bands, [3.6], [4.5], [5.8] and [8.0]. Details of the reduction and catalog preparation can be consulted in \citet{C2d} and through the \anchor{http://ssc.spitzer.caltech.edu/legacy/c2dhistory.html}{SST webpages}.

In addition to the infrared data, we performed a series of point measurements with the Green Bank 70m telescope to determine the variation of the emission of the (1,1) and (2,2) rotational transitions of ammonia (NH$_3$) across the extension of the central core region in B59. The observations were carried out during October 2007 as a complementary project of the ammonia survey of dense cores across the Pipe Nebula by \citet{pipenh3}.

\subsection{Catalog Merging}
\label{section:observations:subsection:catmerge}

The construction of the extinction map requires a catalog that combines the NIR and MIR sets. We joined the photometry catalogs from different observations in a progressive sequence: first, the photometry lists from single filter observations on each instrument were combined into single-field catalogs. Second, the single-field catalogs were merged into single-instrument catalogs. Third, we constructed a complete ESO photometry list, by merging the ISAAC and SOFI catalogs with a source matching algorithm that preferentially listed ISAAC over SOFI detections. Positional matching between NTT and VLT was calculated within a tolerance radius of 0.24$\arcsec$, i.e. twice the average uncertainty of the ESO-NTT astrometric solutions. The final ESO catalog contained 59721 sources. Fourth, the ESO final list was merged to a 2MASS ``bed'' catalog obtained from the All Sky Data Release. The 2MASS bed catalog covers a rectangular area from  $(\alpha,\delta)=(258.30,-27.70)$ to $(\alpha,\delta)=(257.55,-27.55)$, containing 14099 sources after separating those fainter than the listed sensitivity limits and those with photometric quality flags `U' and 'X', which indicate upper-limits and defective observations, respectively. In order to merge our ESO catalog with the 2MASS bed catalog, we determined positional matches in the area within a tolerance radius of 0.36$\arcsec$ (three times the average uncertainty of the ESO-NTT astrometric solutions) and rejected the 2MASS matching sources except in a minority of cases when matching ESO sources were saturated. The total number of sources in the ESO+2MASS catalog is 71216. However, we limited our analysis only to those sources with photometric uncertainty smaller than 0.1 mag in at least one band and no infrared excess (either from young source candidates in the cluster or field OH IR stars, abundant in the region, as shown by \citet{lal06}). Our final NIR working catalog contains a total of 55257 sources.

\begin{figure*}
\begin{center}
\includegraphics[width=7.0in]{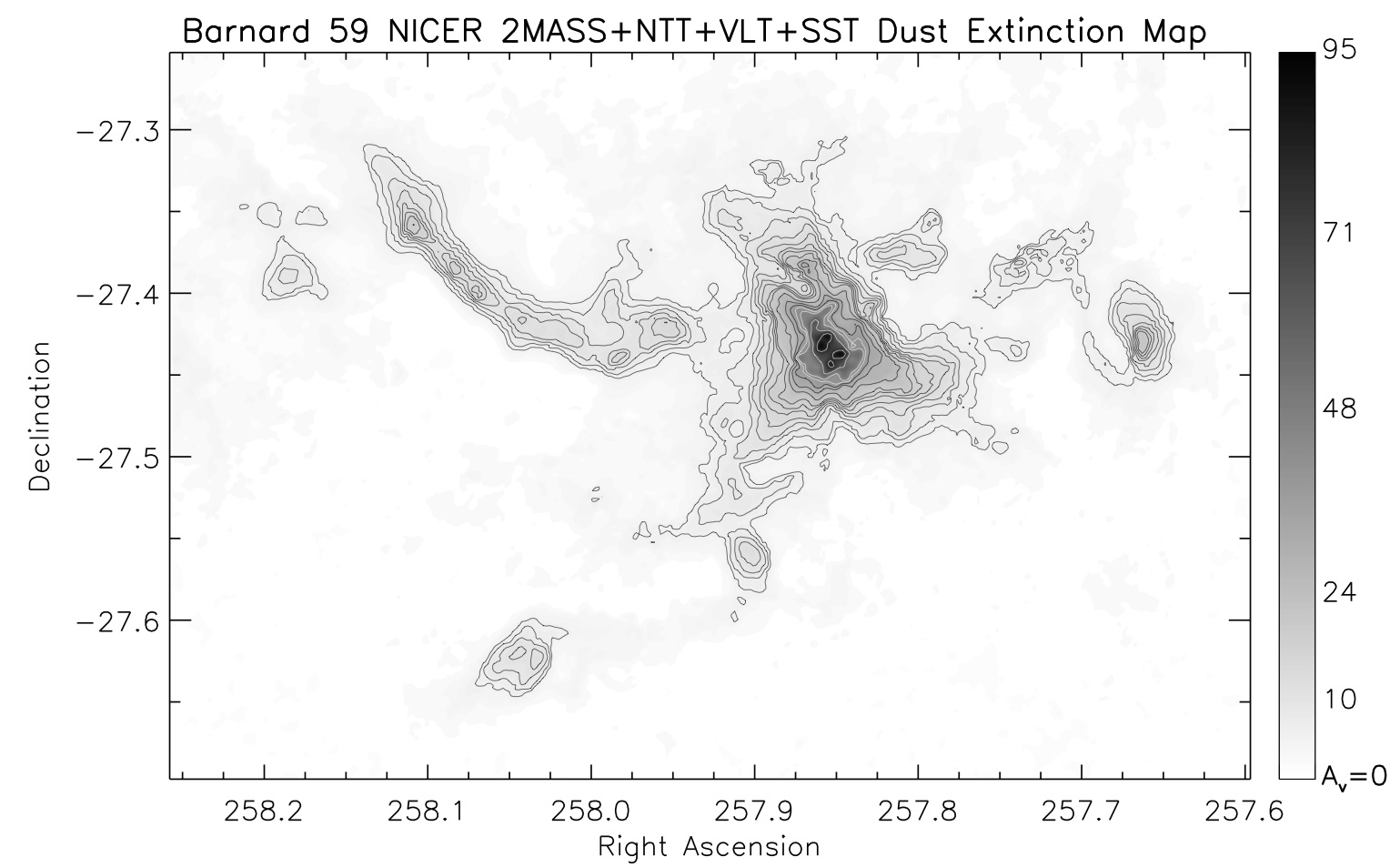}
\caption{Extinction map of the Barnard 59 region at a spatial resolution of 20$\arcsec$. The map is presented as a contour level image with linear increments of 1.0 mag. The black solid line contours mark levels of extinction from $A_V=$6.0 to 20.0 mag in 2.0 mag increments,
and from $A_V$=20.0 to 35.0 mag in 5.0 mag increments; the white contours mark levels of extinction at 40, 45, 60 and 80 mag. At this spatial resolution, the maximum $A_V$ in a pixel is approximately 95 mag, however, measurements on some individual reach up to 100 mag.\label{fig:extmap:all}}
\end{center}
\end{figure*}

Finally, we combined the NIR and the MIR data. In order to do this, we matched positions of sources in the ESO-2MASS catalog with those in the B59 STT C2D public catalog (5400 sources), using a positional tolerance of 0.36$\arcsec$; this way, we found a total of 3683 NIR/MIR counterparts with photometric uncertainty smaller than 0.1 mag in at least one band. However, during the construction of the extinction map, the ESO-2MASS-SST common area catalog was analyzed separately and merged only partially: we only used SST sources within an enclosed region where the density of ESO sources drops dramatically as one reaches the center of the cloud. Within this NIR obscured region, we used a combination of hybrid NIR+MIR and pure MIR colors for sources whose $H$ or $K_s$ detections were null or had unacceptable photometric quality (see section \ref{section:analysis}, below). We also rejected IRAC sources which had C2D flags indicating YSO candidates, stars with dusty envelopes, and stars with flat, rising or cuspy spectral energy distributions (SED). In addition, from the set of objects with available $[5.8]$ and $[8.0]$ photometry, we only kept those with brightness within 9.0 and 14 mag, and with colors $[5.8]-[8.0]<0.5$ mag, a restriction that helps to reject candidate class III sources (their intrinsic values are different from those of background giants) and background galaxies.

\section{An Extinction Map of Barnard 59}
\label{section:analysis}
%\subsection{Application of the Near Infrared Color Excess methods}
%\label{section;analysis:subsection:hnice}

The technique used to construct our extinction map is a combination of the Near Infrared Color Excess (NICE) method originally presented by \citet{NICE} and the revised method (NICER) of \citet{NICER}. The NICER algorithm requires a source to be observed in 3 bands -in our case $J$, $H$, and $K_s$ in order to determine the extinction from two colors, otherwise it uses only two bands and one color, as in NICE. We do not have ESO $J$ band observations available for a majority of the ESO fields; particularly all fields covering the central region. Therefore, for a majority of ESO sources, extinction was determined from $H-K_s$. Also, when IRAC colors were added in the highest column density areas, we found it was more robust to use the color that was available for the largest number of sources possible at a given extinction (see below), rather than choosing two fixed colors from a pool with variable uniformity, hence the use of NICE (i.e. extinction calculated from one color at a time) for MIR and hybrid NIR+MIR colors. Overall, for sources using the NICE method, we used the following general formulation:

\begin{equation}
A_V = C_{el}\times E(\lambda _1 - \lambda _2) = \\
 C_{el}\times [(m_{\lambda _1} - m_{\lambda _2})_{obs} - (m_{\lambda _1} - m_{\lambda _2})_{0}] 
\end{equation}

\noindent where $m_{\lambda _1}$ and $m_{\lambda_2}$ represent the observed magnitude of the source in any two infrared bands. The intrinsic color, $(m_{\lambda _1} - m_{\lambda _2})_{0}$, is calculated in a control field, in our case a low column density area located approximately 2$\arcdeg$ west of B59 at the same galactic latitude. We also have ESO observations and a C2D catalog for this field. $C_{el}$ is a constant derived from the adopted extinction law, in the form:

\begin{equation}
C_{el} = \frac{A_V}{A_{K_s}}{\left(\frac{A_{\lambda _1}}{A_{K_s}} - \frac{A_{\lambda _2}}{A_{K_s}}\right)}^{-1}
\end{equation}

\noindent We used $A_V/A_{K_s}$=0.112 from the extinction law of \citet{riele85}, and the coefficients $A_\lambda/A_{K_s}$ are those from \citeauthor{b59elaw07}. 

For fields where $J$, $H$ and $K_s$ are available (the 2MASS ``bed" catalog described above) $A_V$ can be estimated from two colors, $J-H$ and $H-K_s$, by applying the NICER formulation of \citeauthor{NICER}: supposing that the extinction can estimated as $A_V=a+b1(J-H)+b2(H-K)$ then 
 
\begin{equation}
A_V = b_1[E(J-H)]+b_2[E(H-K)]
\end{equation}

and the coefficients $b_1$ and $b_2$ are calculated through a minimum variance method. In NICER, $A_V$ is required to be an unbiased estimator, i.e. $b1/C_{el,1}+b2/C_{el,2}=1$ and $a+b1(J-H)_0 + b2(H-K)_0=0$.

\begin{figure}
\begin{center}
\includegraphics[width=3.5in]{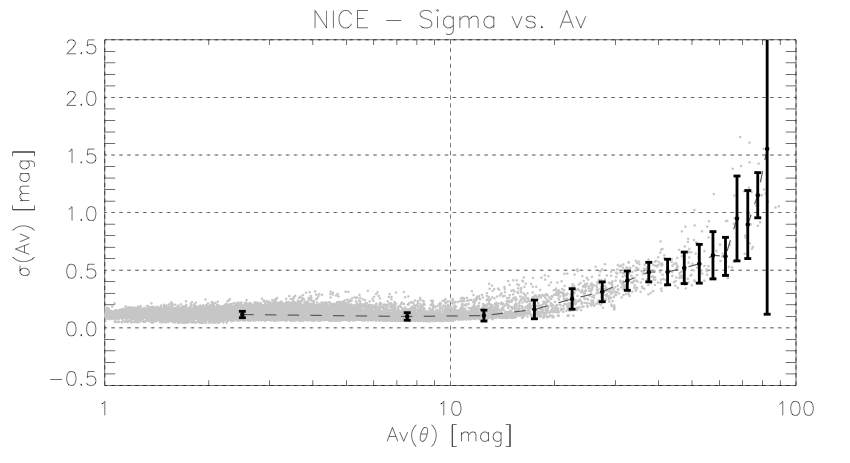}
\caption{$\sigma_{A_V}$ vs $A_V$ distribution for the extinction map of figure \ref{fig:extmap:all}. Individual
measurements per pixel are shown with gray dot symbols. The black dot symbols with error bars joint
by a dashed line indicate the median values in 5 magnitude bins. Notice the
spiking of the noise amplitude for $A_V>60$ mag.  \label{fig:sigmaavp}}
\end{center}
\end{figure}

$A_V$ is evaluated from the $J$, $H$ and $K_s$ observations in most of the survey areas, where column densities were below $\sim4\times 10^{22}\mathrm{cm}^{-2}$ ($A_V\sim40$ mag). At the dense central region of B59, column density rises rapidly to values twice as high. We delimited the highest density region as a rectangular area running from $(\alpha,\delta)=(257.810,-27.470)$ to $(\alpha,\delta)=(257.880,-27.390)$, i.e. slightly larger than the area observed with the VLT. Inside this region we used ESO+IRAC counterparts following a protocol: a) if a source had reliable $H$ and $K_s$ photometry, we evaluated $A_V$ from $E(H-K_s)$; b) if a source was null or unreliable in $H$ then we evaluated $A_V$ from $E([3.6]-K_s)$; c) if a source was null or unreliable in $K_s$ we evaluated $A_V$ from $E([3.6]-[4.5])$; d) finally, if a source was unreliable or null in $[3.6]$ we evaluated $A_V$ from $E([4.5]-[5.8])$. This protocol is summarized in Table \ref{tab:colors}, where we listed the corresponding values of $C_{el}$ and the intrinsic values $(m_{\lambda _1} - m_{\lambda _2})_{0}$ obtained from our control field.

\subsection{Construction of Map} 
\label{section:analysis:subsection:bigmap}

The complete extinction map, presented in Figure \ref{fig:extmap:all} was constructed using a Gaussian spatial filter with FWHM=20$\arcsec$, the beam width was increased to 24$\arcsec$ (20\% oversampling) in the $A_V>40.0$ mag region. The map is Nyquist sampled, yielding pixels separated by 10$\arcsec$. The value of extinction at each position is the Gaussian weighted median of values for individual sources within a radius equal to 3-sigma of the Gaussian beam. The error in the mean extinction, $\sigma _{A_V}$, was calculated as the standard deviation of individual background column density measurements, $\sigma_{*}$, divided by the square root of the number of sources ($\sigma_{*}/\sqrt{N_{*}}$) within the 3-sigma circle. No additional smoothing was applied to the map. In the paper, a simple interpolation routine (\texttt{IDL TRI-SURF}) was used to smooth out contour jagging. The resultant map improves the resolution of the $A_K$ map of \citet{lal06} by a factor of three, allowing us to resolve for the first time: a) the detailed structure of the central core, b) the detailed structure of the thick filament  located  east of the central core (hereafter \textit{eastern filament}), c) the structures of several, peripheral cores located to the southeast and west of the cloud, and d) the faint filaments that connect the main core with the peripheral structures.

\begin{figure}
\begin{center}
\includegraphics[width=3.5in]{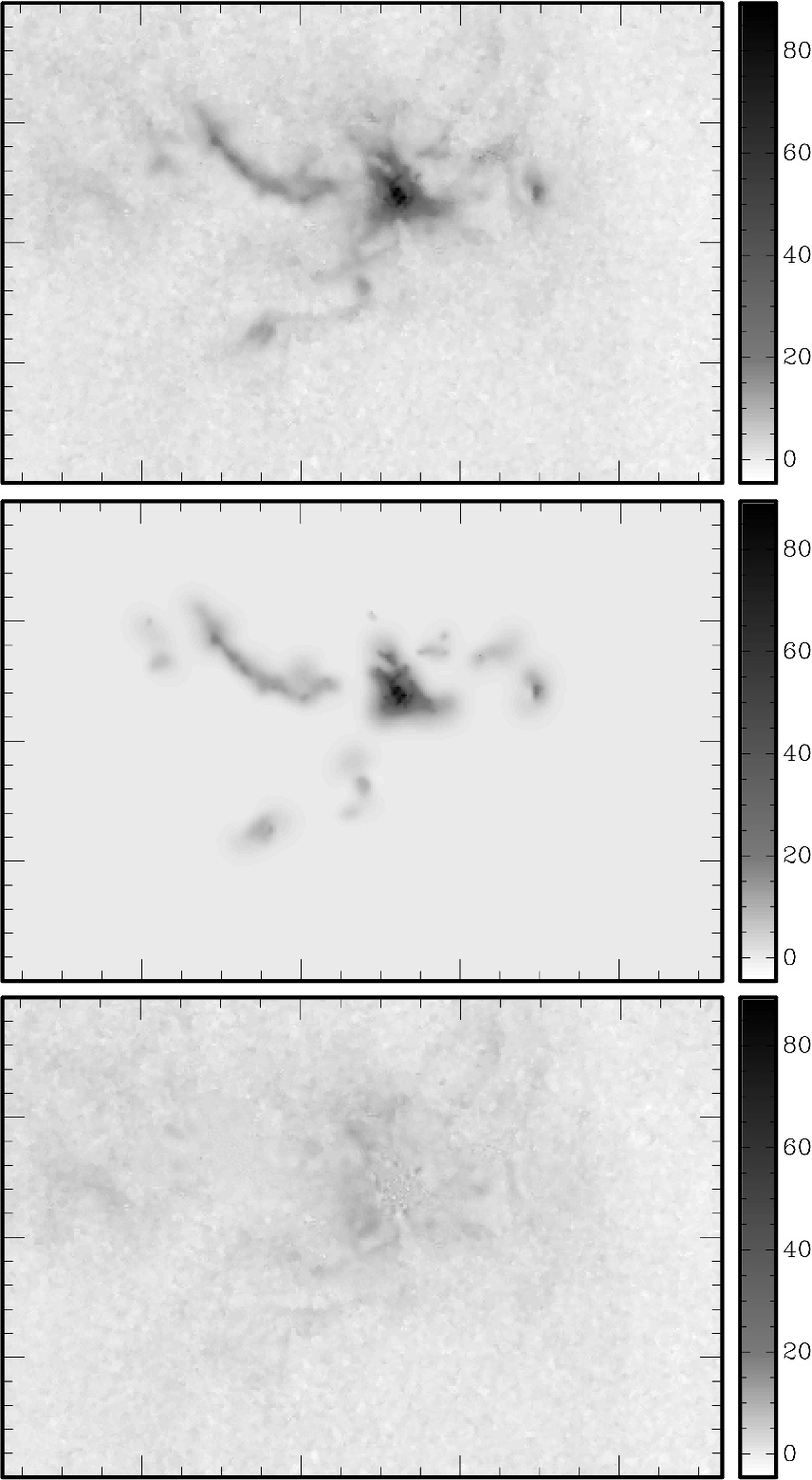}
\caption{An illustrative sequence of the wavelet filter process on the B59 extinction map. From top to
bottom: the original extinction map, a fifth scale wavelet filtered map, and a residual map resulting
from the subtraction of the second from the first. \label{fig:residuals}}
\end{center}
\end{figure}

The spatial resolutions of 20 and 24$\arcsec$ for this whole area map were chosen so that we could assure 1) that the average extinction per pixel could be calculated with at least two stars per pixel all the way to the peak of the central core; 2) that the map reveals most of the features visible at a threshold resolution ($18\arcsec$), at which a number of empty pixels start to appear. Empty pixels would especially affect the central core region, where the number of NIR sources starts to decrease significantly for $A_V>45$.

\subsubsection{Map noise}
\label{section:analysis:subsection:bigmap:subsub:noise}

Assuming that the quality of the photometry is uniform across the map, it is expected that the error in the measurement of extinction at a given line of sight ---i.e. the noise per pixel---, will be inversely proportional to the number of sources involved. In Figure \ref{fig:sigmaavp} we show a plot of the error in the mean extinction, $\sigma_{A_V}$, as a function of $A_V$. In general, at low ($A_V<10$ mag) and moderate (10$<A_V<$30) extinction levels, the error is low, $0.11\pm0.04$ mag, and also uniform, indicating that any structures within those levels are well resolved at the chosen resolution. Above $A_V\approx$30 mag, $\sigma_{A_V}$ increases monotonically until it reaches a maximum median value of about 1.5 mag per pixel within the most obscured regions in the central core ($A_V>$75 mag). At high extinction levels, the significance of features in the map is determined by this monotonic increase of the noise amplitude. For example, a 5$\sigma_{A_V}$ interval in low to moderate extinction regions ($10<A_V<30$ mag) increases slowly from 0.55 to 1.5 mag, but has to be increased from 3.0 to 5.0 mag at the highest extinction levels ($A_V>50$ mag). This indicates that any features in the map, especially peaks and valleys indicating the presence of dense gas cores, are only to be trusted if they can be measured as variations in extinction larger than the intervals just described.

\begin{figure*}
\begin{center}
\includegraphics[width=6.5in]{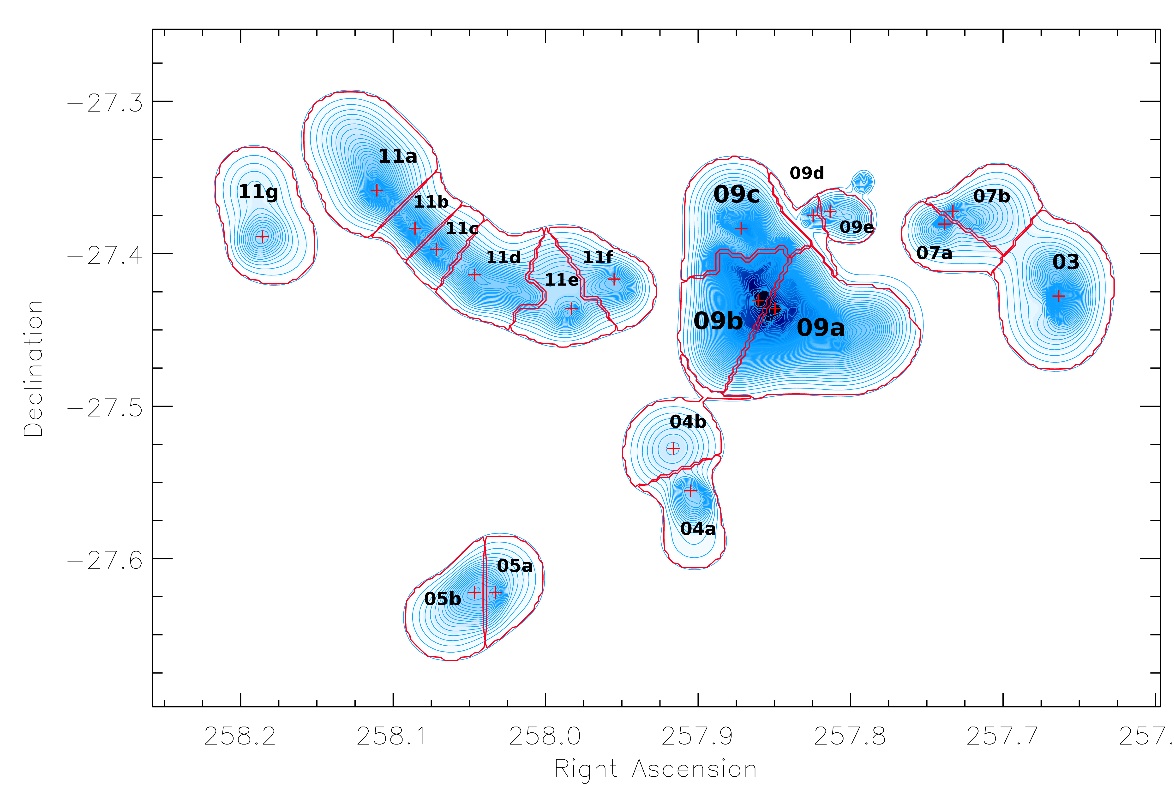}
\caption{Identification of main cloud features in Barnard 59. The map shows the location of cores identified with CLUMPFIND-2D on a wavelet filtered map. The 5$\sigma$ intervals used to identify features in the map as shown as iso-contour levels. \textit{Please see the electronic version of the Astrophysical Journal for a full color version of this figure} \label{fig:identification}}
\end{center}
\end{figure*}

\section{Identification of Dense Cores}
\label{section:identification}

A dust extinction map is a two-dimensional (2D) structure. Potential dense cores can be identified as local maxima, forming extinction peaks that merge with the cloud at a local background level and have roughly elliptical shapes. The identification of individual cloud cores in a 2D map would be relatively straightforward if the extinction peaks were always well separated from each other. In practice, neighboring peaks may be seen in projection close to each other and in some cases overlap, making it difficult to determine their boundaries at the local background level. In order to reduce this effect, we reduced the contribution of the local background by applying a wavelet transform filter to the map using a routine developed by Benoit Vandame (see \citeauthor{alola07}). The routine is based on a Multi-scale Vision Model (MVM), designed to reconstruct astronomical images \citep{mvm97}. The MVM filtering is performed by isolating features rising above a 4.0$\sigma$ threshold in the wavelet transform space at five spatial scales: 0.4, 0.8, 1.6, 3.2 and 6.4$\arcmin$. At each scale, the pixels with values higher than the threshold in the wavelet transform space are identified and labeled, and the program constructs a series of trees of interscale connectivity, which are then used to identify significant objects and to reconstruct the image. The resultant MVM filtered image contains zero or minimum background, making it easier to define the boundaries of individual clumps. A simple check was done by subtracting the filtered image from the original map: the resultant residual image contains only background extinction and low level extinction from a few filamentary structures, mostly located east of the central core (see Figure \ref{fig:residuals}).

\subsection{Significant Features}
\label{section:identification:subsection:significant}

Identification of individual clumps in 2D maps (dust extinction and dust emission maps are common examples) is usually done by means of a clump finding algorithm. 2D maps lack information on the velocity distribution of the gas at each line of sight, making it difficult to separate clumps if they overlap. Fortunately, the case of the Pipe is particularly benign because the cloud is projected across its longest axis on the plane of the sky. Most clump finding algorithms depend on a set of parameters to success at detecting peaks and to define the boundaries of individual features, and thus their reliability is ultimately subject to the correct choice of those parameters. We used the \texttt{CLUMPFIND-2D} algorithm \citep{clumpfind}, because it has only one free parameter, the set of contour levels used to identify adjacent regions associated with a local peak or maxima. The intervals used to define the contour levels are defined by the user and have to be chosen carefully in order to select only significant structure and avoid identification of spurious features \citep{jouni09}. Moreover, as discussed in $\S$\ref{section:identification:subsection:significant:subsection:significance}, additional criteria are required to determine the significance of individual detections.

To define the best set of contour levels, we considered the increase of the noise amplitude as a function of extinction in our map. We chose intervals starting at $A_V=$0.55 mag contour level, equal to 5 times the average noise per pixel in low extinction ($A_V<5$ mag) regions. For low and moderate extinction levels (1$<A_V<10$ and $10<A_V<25$) we used fixed 5$\sigma_{A_V}$ intervals set to 0.55 and 1.25 mag respectively. For thicker regions we calculated the 5$\sigma_{A_V}$ interval corresponding to the increase in the noise as shown in Figure \ref{fig:sigmaavp}, until reaching $A_V$=73 mag. In a first run, we kept a list of all cores detected with the chosen set, then we decreased the level values by 1 and 2 sigmas (test level dropping) in order to discard spurious detections. A few features were no longer identified after the level dropping and were considered as spurious detection. Also, we increased the minimum number of pixels for region identification from 20 to 50, to avoid confounding single pixel fluctuations with extinction peaks. Our final \texttt{CLUMPFIND-2D} run yield 20 significant extinction peaks in the map, as shown in Figure \ref{fig:identification}. A list of these peak identifications is presented in Table \ref{tab:clf2d}. 

\subsection{Feature Properties}
\label{section:identification:subsection:significant:subsection:properties}

For each of the extinction peaks identified, \texttt{CLUMPFIND-2D} defines an optimized boundary at the bottom contour level (0.55 mag, in our case). The size of a core or peak gets defined as the equivalent radius of the boundary. Mass was estimated by summing the background corrected total extinction in pixels within the boundary. The conversion to mass, assuming a standard value for the gas to dust ratio $N_H/A_V = 2.0\times10^{21}\mathrm{cm}^{-2}$, is given as:

\begin{equation}
M_{core} = 1.28^{-10}\left (\frac{\theta}{\arcsec}\right )^2 \left(\frac{D_{cloud}}{\mathrm{pc}}\right)^2 \sum_{i=1}^{N}{f_i(A_V)}\, M_\odot
\end{equation}

\noindent Where $\sum_{i=1}^{N}{f_i(A_V)}$ adds the contribution of $N$ pixels within the feature boundary in the wavelet filtered map, $\theta$ is the beam size, and $D_{cloud}$ is the distance to the cloud. In our case, $D=$130 pc and $\theta =20\arcsec$.

Using the values for mass and equivalent radii, we made estimations of the average density, $\bar{n}$ of the features and, assuming a gas temperature of 10 K we calculated the Jeans length and the Jeans mass for each feature as:

\begin{equation}
L_J = 0.2\left (\frac{T}{10\mathrm{K}}\right )^{1/2} \left ( \frac{\bar{n}}{10^4\mathrm{\ cm}^-3}\right )^{-1/2} \mathrm{\ [pc]}
\end{equation}

and

\begin{equation}
M_J = 1.88\left (\frac{T}{10\mathrm{K}}\right )^{3/2} \left ( \frac{\bar{n}}{10^4\mathrm{\ cm}^-3}\right )^{-1/2} \mathrm{\ [M}_\odot\mathrm{]}.
\end{equation}

\subsection{Significance of features}
\label{section:identification:subsection:significant:subsection:significance}

The next task was to determine which of the features we identify correspond to cloud cores previously identified in the study of RLA09. The core list of RLA09 for the same region contains only 7 cores, resulting of merging some of the individual features they identified in the map of LAL06. The merging of features in RLA09 was done by following a well defined prescription: a feature is significant if its peak rises by at least 3$\sigma$ above the noise, and two features represent independent cloud cores only if their separation is larger than the mean Jeans length of the cloud (approximately 0.26 pc) \textit{or} if their central velocities (measured from C$^{18}$O observations) differ by more than the 1-D isothermal sound speed at $T=10$ K.

\begin{figure}
\begin{center}

\includegraphics[width=3.5in]{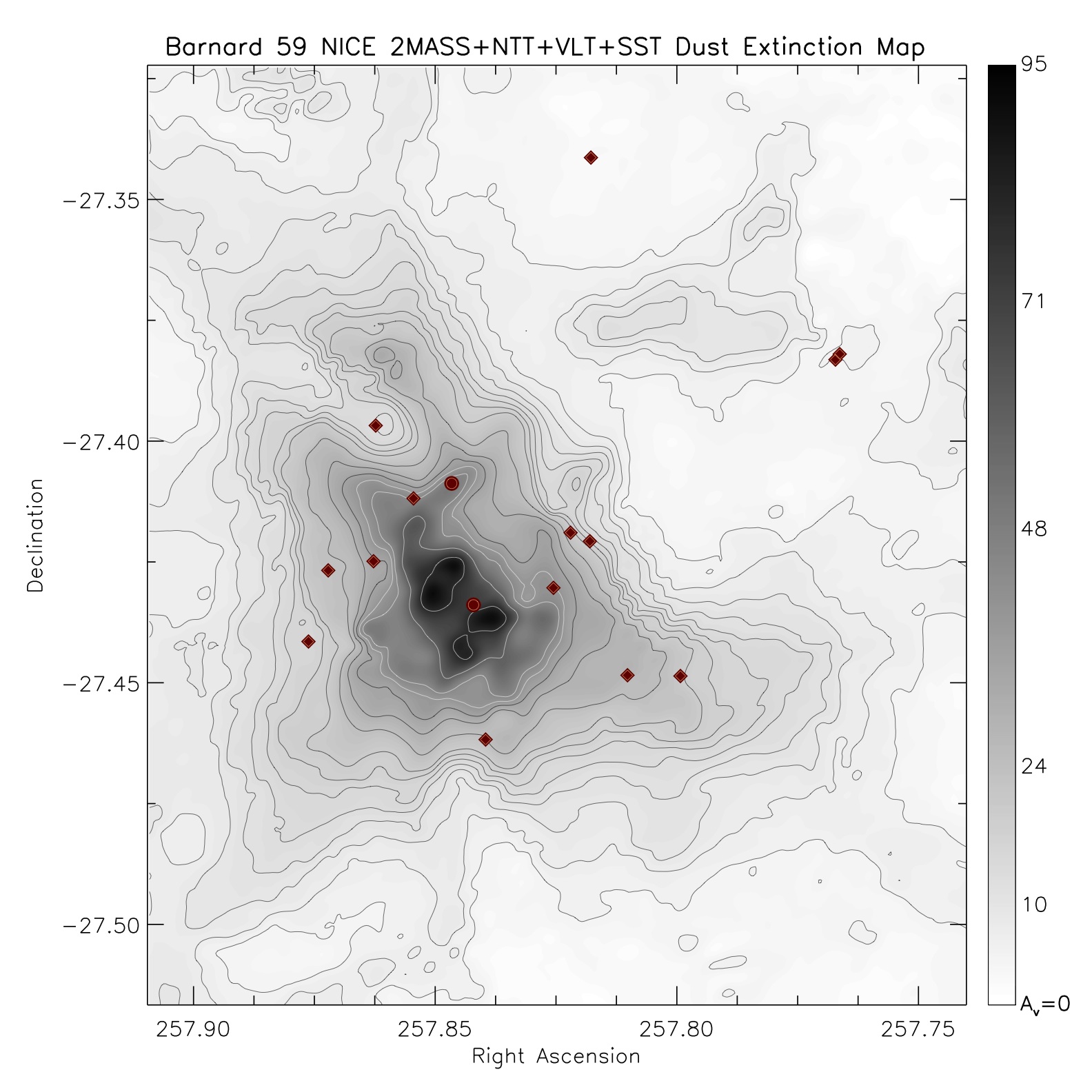}
\caption{A close look map of the central core in B59, at the same resolution as the map in Figure \ref{fig:extmap:all}. The diamond shaped symbols indicate the locations of YSO candidates identified by Brooke et al. (2007). The circle shaped symbols indicate sources identified as Class 0/I.\textit{Please see the electronic version of the Astrophysical Journal for a full color version of this figure} \label{fig:B59C}}
\end{center}
\end{figure}

Each of the 20 extinction peaks identified in our map, coincide with at least one of the 7 cores defined by RLA09. Therefore, many of the features in our map may not be independent cores but they may represent significant {\it substructure} within larger, independent cores. 

Our high resolution extinction map allows us to measure improved masses and sizes of features identified with \texttt{CLUMPFIND-2D} and therefore provides improved measurements of the \textit{local} Jeans length (LJL) in the region, that we can then compare to the peak to peak projected separations of features which are also slightly better defined in our map. In what follows, we will discuss the significance of the features found in our map and, we will make a distinction between substructure peaks and independent cores based on the rules of RLA09. We cannot apply the rule of the difference of central velocities because we do not have pointed radio observations for each of the peaks we identified, but we can use the projected peak to peak separations in our map to determine if any individual extinction peaks are separated by more than the LJL. 

\subsubsection{Central Core: structure and relation to young cluster members}
\label{section:identification:subsection:significance:subsub:centralcore}

In Figure \ref{fig:B59C} we show a close-up map of the central core region. We also marked the locations of the YSO candidates identified by BHB07\footnote{Additional information as well as list of ESO deep level photometry and a color-magnitude diagram for the cluster candidates is presented in a subsequent spectroscopic study of the Barnard 59 cluster (Covey et al., in preparation)}. We confirm that 13 of the 20 YSO candidates are located within the central core, specifically above extinction levels of $A_V=15$ mag, and out of these, 12 are located below $A_V=50$ mag. Only one star, a class 0/I source (source No. 10 in BHB07 list) is projected towards an extinction higher than $A_V=70$ mag.

The central core is approximately trapezoidal in shape; the Northwest corner of the trapezoid appears to stand out from the main body at $\delta=-27.40$, defining the core 09c. This core is significant, with at least three closed contours at the 5$\sigma_{A_V}$ level between $A_V$=25 and 30 mag. Moreover, the projected distance from 09c to the midpoint between 09a and 09b, is about 0.12 pc, slightly longer than the LJL at the central region. Core 09c is located next to what seems to be a cavity opening around $(\alpha,\delta)=(257.86, -27.40)$ --coincident with source 13 in BHB07 list. The cavity in the cloud is likely to be related to the gas outflow, evidenced by the ample line wings of $^{12}$CO lines \citep{onishi99,lohr07}. The most likely candidate responsible for excavating the cloud is source 2M171123 \citep{riaz09}, listed as Source No. 11 in BHB07 list, also associated with the millimeter emission source B59-MMS1 \citep{reipurthb5996}. In the IRAC images, a significant scatter light feature appears to confirm the action of the source on the surrounding material. 

\begin{figure}
\begin{center}

\includegraphics[width=3.5in]{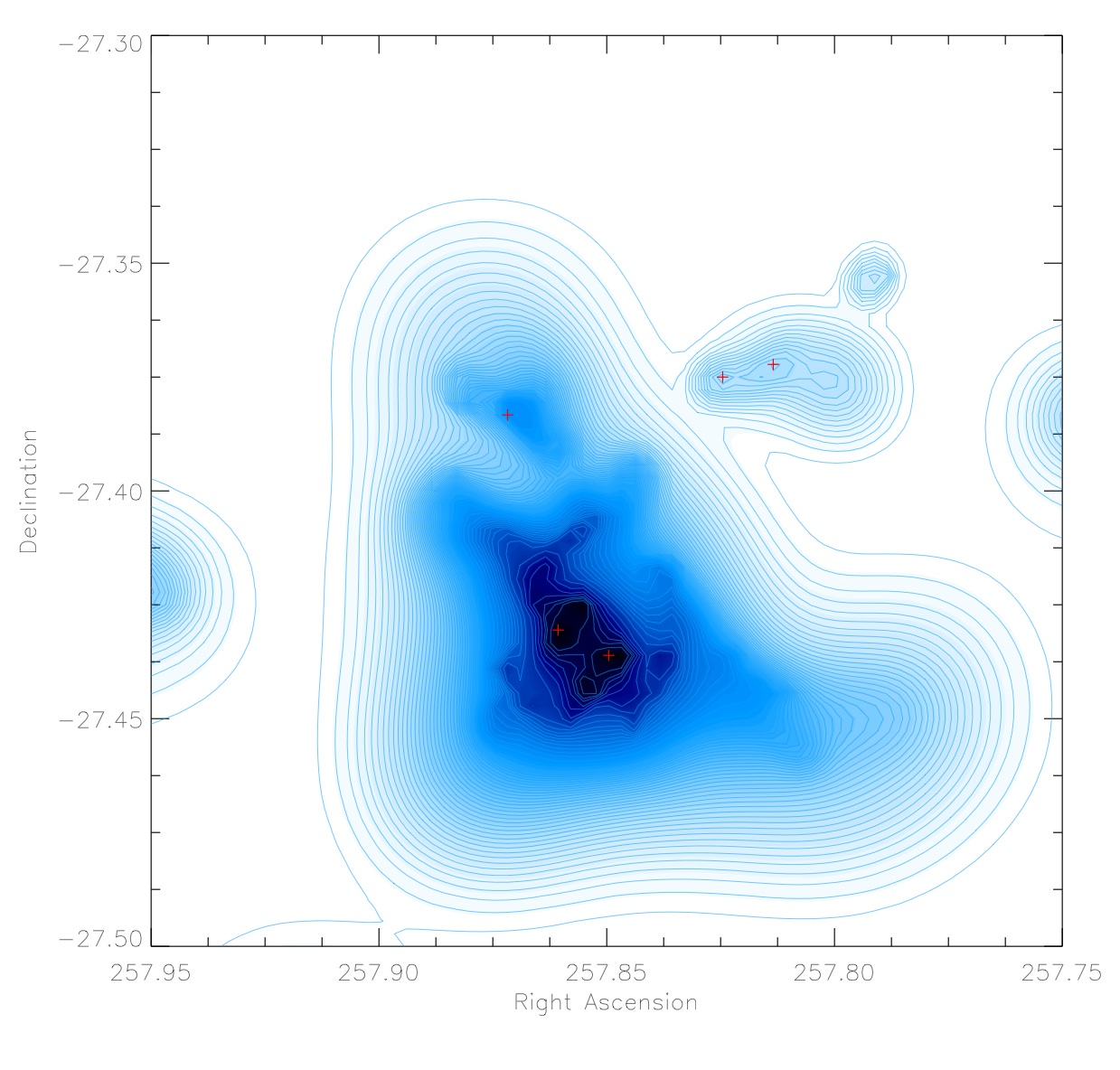}
\caption{A close look of the central core region in the identification map of figure \ref{fig:identification}. The 5$\sigma$ intervals used to identify features in the map as shown as iso-contour levels. From top to bottom, the five top levels are $A_V$=73,66,60,55 and 50 mag. \textit{Please see the electronic version of the Astrophysical Journal for a full color version of this figure} \label{fig:09clf}}
\end{center}
\end{figure}

Our map appears to show a second indentation in the core at $(\alpha,\delta)=(257.842, -27.478)$, just south of source No. 9 in BHB07 list (2M171121), which also shows a significant scatter light feature in NIR images and it is possibly associated with an outflow structure \citep{riaz09}. If so, the indentation in the map could be an indication of another cavity like structure being formed by a protostellar outflow. Outflows from young stars have been long known to play an important role in the evolution of star forming clouds \citep[e.g. ][]{circinus99, matzner00}. Fragmentation enhanced by outflows has also been predicted by recent numerical studies \citep{nakali07}. 

Regarding the central region, our map reveals structural details that could not be previously resolved in lower resolution maps. In Figure \ref{fig:09clf} we show a ``close up" of the identification map with the 5$\sigma_{A_V}$ intervals used for \texttt{CLUMPFIND-2D}. Starting from the top, at the $A_V=$73 mag contour, the core appears to split in two with source No. 10 located in the middle. The central split is significant at a 5$\sigma_{A_V}$ level at the maximum local noise amplitude ($2.0<\langle\sigma_{A_V}\rangle<3.0$ if $65<A_V<80$ mag). At the next contour level, $A_V$=66, the contours close into a single feature, and no more substructure is found until we go down to the $40<A_V<50$ mag range. There, the map shows some small closed contours, but we found that they do not represent significant features, as they cannot longer be detected if the levels are dropped by 1$\sigma$. 

In summary, \texttt{CLUMPFIND-2D} suggests that the B59 core might be split into three fragments: two defined by the split in the filtered map at $A_V$=73 mag (listed as 09a and 09b in the identification list of Table \ref{tab:clf2d}), and one near the northern cavity (listed as 09c in the identification list). However, in section \ref{section:discussion:subsection:centralcorefrag:nosubstructure} we argue that the split near the center might not be as physically significant as \texttt{CLUMPFIND-2D} suggests, and that the central core is likely a single stratified core with only one independent new fragment (09c) identified in this high resolution map.

\subsubsection{Peripheral Cores}
\label{section:identification:subsection:significance:subsub:roundcores}

Four cores are located in the periphery of the complex, corresponding to cores 3, 4, 5 and 13 in the list of RLA09. These objects have a morphology reminiscent of Bok globules, but in our map, they appear to be still connected to the rest of the cloud region by thin ($3<A_V<6$ mag) filaments. Core 03 is clearly centrally condensed, with an elongated N-S distribution near the center; it also has a very extended ``tongue" at its northern edge which makes it 40$\arcsec$ wider at the top; its peak extinction is 19.9 mag in the filtered map. Core 05 has a peak extinction of 10.9 mag in the filtered map and has a toroidal like structure, similar to Globule 2 in the Coalsack cloud \citep{coalsack}, which \texttt{CLUMPFIND-2D} picks as a double core. We merged the two fragments, 05a and 05b because their projected separation is shorter than their LJL. In the case of peaks 04a and 04b, their morphology in the map and optical images might suggest they are independent, but as their projected separation (0.10 pc) is about half their LJL ($\sim$0.2 pc), we cannot tell if they are independent. Thus, we merged features 04a and 04b into a single core, 04ab.

There are two highly elongated clumps at the Northwestern edge of the B59 region, both significantly larger in the E-W direction than in the N-S direction. \texttt{CLUMPFIND-2D} suggests these cores are each divided in two fragments (cores 09d,e and 07a,b). The peak-to-peak projected distance from 09d to 07a is about 270$\arcsec$ or 0.17 pc, similar to the LJL of both cores, thus we confirm that peaks 09 and 07 are independent. 

In the case of peaks 07a and 07b we find that the projected distance (0.06 pc) is shorter than the LJL, thus we merged into a single core, 07. Same criterion is applied to peaks 09d and 09e, which we merged into core 09de.

%\begin{rotate}
\begin{deluxetable*}{lcrrcrccc}
\tablecolumns{9}
\tablewidth{0pc}
\tablecaption{\texttt{CLUMPFIND-2D} Identifications in high-resolution extinction map of Barnard 59 \label{tab:clf2d}} 
\tablehead{
\colhead{Clump ID\tablenotemark{1}} &
\colhead{Peak $A_V$} & 
\colhead{Sum $A_V$} &
\colhead{No. pixels} &
\colhead{$R_{eq}$ (pc)} &
\colhead{Mass} &
\colhead{$\bar{n}$ } &
\colhead{$L_J$} &
\colhead{$M_J$} \\
\cline{2-3}
\colhead{} &
\multicolumn{2}{c}{[mag]} &
\colhead{in map} &
\colhead{[pc]} &
\colhead{[M$_\odot$]} &
\colhead{10$^4$ cm$^{-3}$]} &
\colhead{[pc]} &
\colhead{[M$_\odot$]} \\
}
\startdata 

\multicolumn{9}{c}{Central Core Region}\\*
\cline{1-9}\\*	

 09a & 83.3 &  11995.8 &    1042 &   0.11  &  10.4  &  2.87  &   0.12  &   1.1\\
 09b & 83.7 &	9796.4 &     530 &   0.08  &   8.5  &  6.46  &   0.08  &   0.7\\
 09c & 24.0 &	3254.0 &     554 &   0.08  &   2.8  &  2.00  &   0.11  &   1.1\\
 09d &  7.8 &	 528.7 &     209 &   0.05  &   0.5  &  1.41  &   0.17  &   1.6\\
 09e &  7.8 &	 188.6 &     105 &   0.04  &   0.2  &  1.41  &   0.17  &   1.6\\

\multicolumn{9}{c}{Eastern Filament}\\*
\cline{1-9}\\*	

 11a & 17.2 &   2954.7 &   760 &  0.10  &   2.6  &  1.20  &   0.18  &   1.7\\
 11b & 14.9 &	 926.6 &   197 &  0.05  &   0.8  &  2.69  &   0.12  &   1.1\\
 11c & 14.5 &	 831.4 &   171 &  0.05  &   0.7  &  2.99  &   0.12  &   1.1\\
 11d & 11.5 &	2017.4 &   417 &  0.07  &   1.7  &  1.91  &   0.14  &   1.4\\
 11e & 11.0 &	1492.9 &   347 &  0.07  &   1.3  &  1.86  &   0.15  &   1.4\\
 11f & 11.3 &	1181.4 &   330 &  0.07  &   1.0  &  1.59  &   0.16  &   1.5\\
 11g &  7.8 &	1741.4 &   613 &  0.09  &   1.5  &  0.93  &   0.17  &   1.9\\

\multicolumn{9}{c}{Peripheral cores}\\*
\cline{1-9}\\*	

 03   & 19.9 &	2747.7 &   811 &  0.10  &   2.4  &  0.96  &   0.20  &   1.9\\
 04a  &  9.4 &   887.7 &   342 &  0.07  &   0.8  &  1.13  &   0.19  &   1.8\\
 04b  &  4.6 &   778.1 &   393 &  0.07  &   0.7  &  0.81  &   0.22  &   2.1\\
 05a  & 10.9 &	 804.8 &   285 &  0.06  &   0.7  &  1.35  &   0.17  &   1.6\\
 05b  &  9.9 &  1301.0 &   396 &  0.07  &   1.1  &  1.32  &   0.17  &   1.6\\
 07a  & 14.2 &	 707.5 &   259 &  0.06  &   0.6  &  1.36  &   0.17  &   1.6\\
 07b  & 10.3 &  1079.4 &   339 &  0.07  &   0.9  &  1.39  &   0.17  &   1.6\\
\enddata
\tablenotetext{1}{Core identifications follow the numbering used in the 
list of RLA09. Identified substructure is named after the main RLA09 core number plus
letters (a,b,c, etc.).}

\end{deluxetable*}
%\end{rotate}

In the case of peaks 09d and e, the analysis of RLA09 suggested they share the same central velocity value as the central core, 09a-c, and for that reason they were merged to it. However, in our map, the projected distance from the midpoint between 09d and 09e to the midpoint between 09a and 09b is 0.16 pc, which is comparable to the LJL and larger than the LJL of 09a and 09b. Thus, we consider peak 09de to be independent from the central core. 

In RLA09, peak 11g was merged to core 11 (the eastern filament) as it was found to have the same central velocity. However, we find that the gap between peaks 11g and 11a in the map is very significant (in optical and NIR images the gap is transparent to background stars), and that the projected separation from peak to peak, 0.18 pc, is about the same as the LJL in both cores (0.17-0.18 pc). Thus, we suggest that 11g is an independent core.  

%\subsubsection{Elongated Cores and Eastern Filament}
%\label{section:identification:subsection:significance:subsub:filaments}

The eastern filament in B59 extends east to west from $(\alpha,\delta)=(257.9377,\ -27.4193)$ to $(\alpha,\delta)=(258.1793,\ -27.3113)$. The length from tip to tip, as projected on the map, is approx. 0.57 pc. Its easternmost tip, listed as 11a (core 11 in the list of RLA09) has a peak extinction of 20 mag. We identified 6 extinction sub-peaks across the filament, which we find to be suggestive of significant sub-structure, as we discuss in $\S$\ref{section:discussion:subsection:grandfilament}.

\section{A Brief Discussion on Fragmentation in B59}
\label{section:discussion}

The high spatial resolution achieved in our map should allow us to determine the presence of substructure related to fragmentation in the central core of the Barnard 59 complex. However, the map does not show compelling evidence of multiple fragmentation, which might be expected from a core that has already formed more than a dozen stars and has been active for at least the age of the stars, estimated to be 2$\pm$1 Myr (Covey et al 2009, in preparation).

\subsection{Fragmentation of the Central Core: A Lack of Substructure}
\label{section:discussion:subsection:centralcorefrag:nosubstructure}

As described in $\S$\ref{section:identification:subsection:significance:subsub:centralcore}, \texttt{CLUMPFIND-2D} divides the region below the cavity in the central core in two pieces given that there are two closed contours at the $A_V$=72 mag, the highest contour level we used. Put together, the fragments have a total mass of 19 $M_\odot$, which is about 10 times as large as the critical Bonnor-Ebert mass of the cloud, 2 $M_\odot$ \citep{pipecores} and almost 20 times as large as the local Jeans mass, estimated to be close to 1 $M_\odot$. However it is not clear whether the split might represent the existence of independent cores: the split at the $A_V$=72 mag level is significant above the local noise amplitude, but as we move away from this central region towards lower column density values, we cannot find any other evidence of significant substructure. The total extinction above the second highest contour level at $A_V$=66, is 2422 mag, distributed over 33 pixels, and yielding a mass of approximately 2.1 $M_\odot$ --that is, if we include the column density in those pixels all the way to $A_V$=1 mag. However, if we count the extinction using a baseline level of $A_V$=66 mag, the two peaks at $A_V$=73 would represent two relatively small fluctuations above a smooth overall density gradient in a single stratified massive core. In such case, the extinction above $A_V$=73 mag would be equivalent to 0.55 M$_\odot$, or about 3\% of the total mass in the core 09ab. 

The cores 09a and 09b have a peak to peak projected separation of 50$\arcsec$ or $\sim$0.03 pc, which is small compared to our estimate for the Jeans length at the two peaks (0.11 and 0.08 pc for cores 09a and 09b, respectively). Following the prescription of RLA09, the only other possibility for the two fragments to represent independent cores would be that they are separated in velocity by more than the local sound speed. A high resolution C$^{18}$O(2-1) map of the central region in B59, recently obtained at the IRAM 30m telescope (Rom\'an-Z\'u\~niga et al., in preparation) allowed us to determine that the central velocities at the two peaks barely differ. Our NH$_3$ single point measurements also do not show significant variations of the central velocities near the two peaks. Thus, given our current definitions, the two peaks at the nucleus of B59 would be considered to be part of the same core. Moreover, the dust column density distribution is in fact consistent with a single, centrally condensed, stratified core with two small fluctuations near the top. It is significant that the depression in between the two peaks roughly coincides with the location of the protostellar source No. 10. This could suggest that the double peak near the center of the map could be a consequence of the presence of the protostar. 

\begin{figure}
\begin{center}
\includegraphics[width=3.5in]{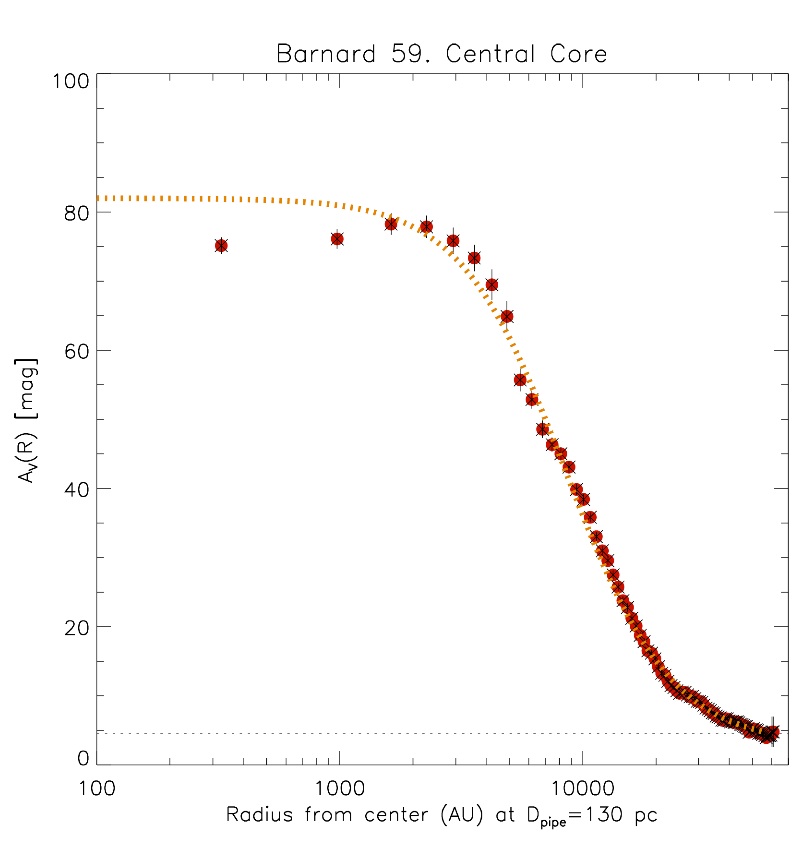}
\caption{Radial profile of the central core 09ab (data points), compared to a King density profile model (thick, dotted curve). The center of the core was defined as the mid-point between cores 09a and 09b. \textit{Please see the electronic version of the Astrophysical Journal for a full color version of this figure} \label{fig:rpk2}}
\end{center}
\end{figure}

If 09ab was a single, large core of 19 M$_\odot$ in the process of collapse, it would be expected to produce, at the current star formation efficiency of 20\%, a star with a mass of approximately 4 M$_\odot$, which would be the most massive star produced in the B59 at the current stage of evolution. To test the idea that the central core could be a single, centrally condensed, quasi-isothermal structure, we compared the radial profile of the 09ab region with an analytical King density profile model \citep{king62} of the form $Y(r)=Y_0[(1+(r/r_0)^2]^{-1}$, as shown in Figure \ref{fig:rpk2}. The fit is done in the $A_V$ vs. radius space with units mag and AU, respectively. The radial profile of the B59-09ab core was constructed with circular, concentric apertures, Nyquist sampled and centered in the mid point between the two peaks visible at $A_V$=72 mag. For that reason, the profile is depressed near the center. Our model of choice fits the core to its peak value, with $Y_0=83$ mag and $r_0=0.85e4$ AU. The reduced $\chi ^2$ value for the fit is 1.81, which represents a moderately good fit, and if we compare only the values for $r>1000$ AU then the reduced $\chi ^2$ falls short of 0.97 indicating a good quality fit. The agreement of the model with our data is thus acceptable, with only the two centermost points deviating significantly from the fit, showing a depression. This depression near the center of the core, accounts for only a 4\% difference in mass with respect to the area under the model profile. The agreement of the data points with the King model reinforces the idea that the core could be a single, centrally condensed, nearly isothermal structure. At this point, we did not want to compare the profile of B59 with a Bonnor-Ebert sphere model as in \citet{b68}. The main reason is that preliminary analysis yield a fit with a very large $\xi _{max}$, indicative of a highly condensed object far from stable equilibrium, and as shown by \cite{kandori05}, it is complicated to distinguish a Bonnor-Ebert sphere in unstable equilibrium from an object in stage of collapse. Such discussion has important consequences, but they fall a bit far from the scope of this study. In a companion paper (Rom\'an-Z\'u\~niga et al., in preparation), we compare the profiles of cores across the whole Pipe Nebula, including B59, with models of singular isothermal spheres, and discuss the implications of such analysis.

\subsection{A Quiescent Cluster Forming Core?}
\label{section:discussion:subsection:centralcorefrag:quiescent}

We used our NH$_3$ emission measurements to obtain information about the kinematic and thermal properties of the dense gas in the center of B59. As expected for a high density core, the NH$_3$ emission was found to be relatively intense for most of the pointed observations. The GBT was pointed towards a total of 37 positions across the central core; 26 of these pointed observations yield significant emission in both the (1,1) and (2,2) transitions, while the 11 remaining pointings were rejected for having a low signal-to-noise ratio in either the (2,2) or the (1,1) transition. The 26 final pointings allowed to determine velocity dispersions, $\sigma _v$, and kinetic temperatures, $T_k$, across the central part of the core via a forward-fitting routine (for details on the data reduction and model fit methods, please see \citet{pipenh3}). The kinetic temperatures range from 10 to 13 K, with a mean of 11.3$\pm0.7$, and the velocity dispersions, $\sigma _v$ varied from 0.13 to 0.29 with a mean of 0.18$\pm$0.05.

\begin{figure}
\begin{center}
\includegraphics[width=3.5in]{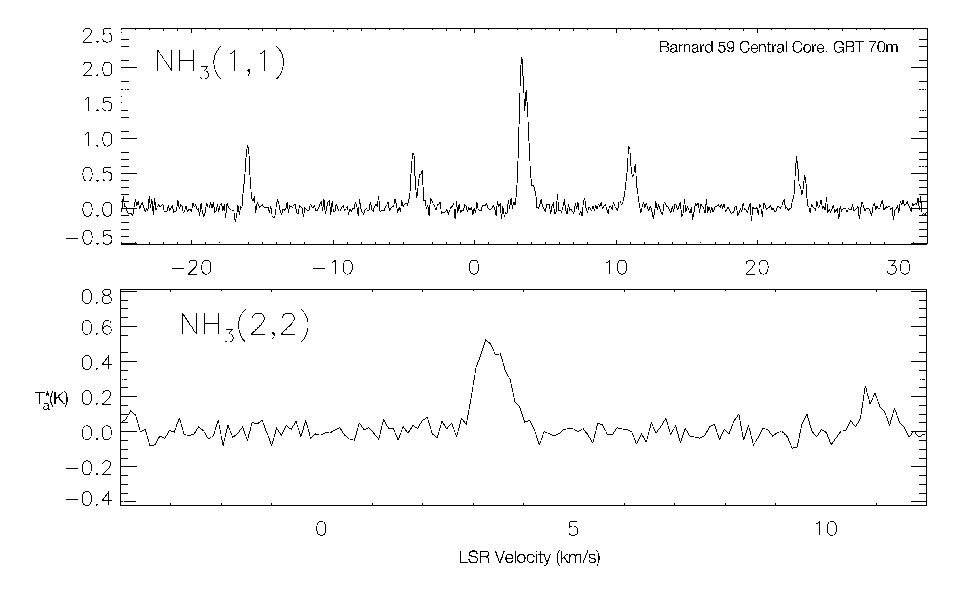}
\includegraphics[width=3.5in]{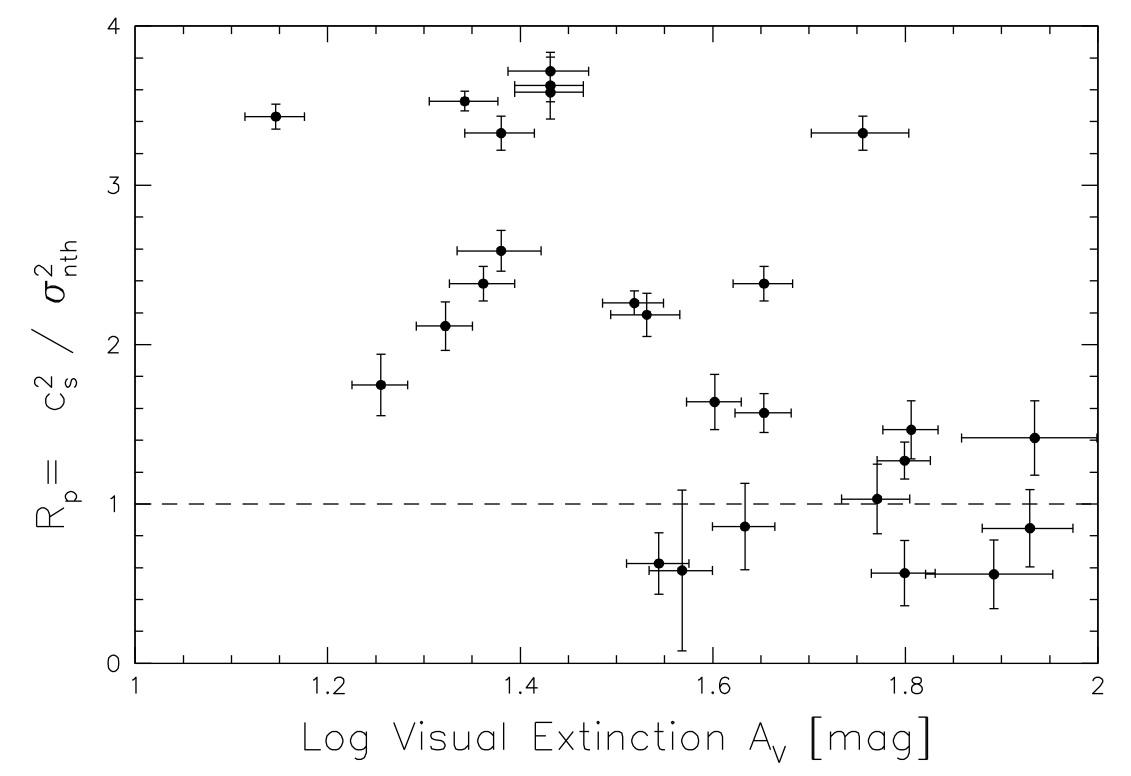}
\caption{Top panel: average spectra of NH$_3$ (1,1) and (2,2) emission towards the center of B59. Bottom panel: Thermal to non-thermal pressure ratio vs. visual extinction in B59. \label{fig:rav}}
\end{center}
\end{figure}

From these measurements we were able to estimate the isothermal sound speed, $c_s$ and the non-thermal velocity dispersion, $\sigma_{nt}$. Then, we calculated the ratio of thermal to non-thermal pressure $R_p=c_s^2/\sigma^2_{nt}$, ($c_s$ is the isothermal sound speed at the corresponding $T_K$), and how it varies across the center of the core. In Figure \ref{fig:rav} we plot the ratio $R_p$ vs the extinction value at the center of the pointing. Errors were estimated as follows: in the case of $R_p$ we took into consideration the uncertainties of the forward fitting routine for $T_K$ and $\sigma_v$; in the case of $A_V$ we considered both the value of the median noise at the value of the pixel closest to each and the mean absolute deviation of the $A_V$ values in the 9 surrounding pixels to account for the uncertainty due to the differences in beam sizes (GBT:30$\arcsec$ vs. NICER map:24$\arcsec$).  

It is certainly striking that a majority of the values are greater than $R_p$=1, indicating that thermal motions dominate over non-thermal (turbulent) motions of the gas in the central core. The mean of $R_p$ in B59 is 1.90$\pm$0.99; for comparison, a thermally dominated core like FeSt-1457 (which also belongs to the Pipe Nebula), was shown to have an average ratio of $\bar{R}_p=3.75\pm1.95$ \citep{aguti}. The plot of Figure \ref{fig:rav} also shows that the 12 lowest values of $R_p$ --with a mean of $R_p=1.0\pm0.4$-- are at the densest part of the core ($\log{A_V}>$1.5 mag). This suggests that non-thermal motions make a more significant contribution to the internal pressure at the central part of the core, where the column density is higher. The relatively large $R_p$ ratios in B59 tell us that the central core is still thermally dominated despite the presence of 14 cluster members packed in the same observed region. Moreover, most of the non-thermal velocity dispersion values in B59 are smaller than the corresponding isothermal sound speed, suggesting that this cluster forming core is characterized by sub-sonic turbulent motions. 

\subsection{Implications}
\label{section:discussion:subsection:centralcorefrag:implications}

Our observations appear to indicate that a) there is a lack of significant sub-structure in the central region of B59 and b) that the central core is mostly quiescent, and still thermally dominated. The result is somewhat unanticipated, as we expected to find evidence of multiple fragmentation in a core with $M\sim 20 M_J$. Moreover, the significant sub-structure we do detect in the central core of the B59 system appears to be related to the embedded protostars, while the overall radial profile of the core seems to be consistent with one large, isothermal structure. Therefore, the question that arises is: what prevents B59 from fragmenting and collapsing into new stars after a period considerably larger than its own dynamical timescale? In one theoretical scenario, a first generation of stars is formed, and they provide feedback which would contribute to suppress fragmentation in the remaining gas, allowing for accretion of material to form a more massive central object\citep{mckeetan02,krumtan07}. Recent observations show that the addition of significant non-thermal motions could prevent fragmentation and allow for the aggregation of larger amounts of material that could later collapse into a massive star \citep{zhang09}. 

As we discussed before, our map shows three possible signatures of stellar feedback in B59: the first is the large northwestern cavity being created by source No. 11, which might have triggered the formation of core 09c; the second is the other cavity-like indentation at the southeast edge of the central core, possibly associated with source No. 9; the third is the depression at the center of the core, associated with source No. 10. However, we cannot detect significant non-thermal motions produced by the stellar feedback in the linewidths of the radio observations. Therefore, additional non-thermal support from some other process might be required.

One possibility is that additional support against collapse could be provided by a mechanism of ambipolar diffusion via a strong magnetic field. \citet{kf04} showed that massive molecular cores tend to have larger column densities than expected from pure thermal support, but also that in some cases the non-thermal linewidths are smaller than expected (as in B59), in which case, support from a strong magnetic field would be required. The hypothesis of a strong magnetic field acting in the Pipe Nebula has been supported by numerical simulations \citep{nakali08}, which show that ambipolar diffusion is an agent capable of slowing down the process of condensation, as needed in the Pipe, where the free-fall time scale is very short given the median core density. A magnetic field could provide the necessary non-thermal support against collapse in massive cores like B59, giving enough time for turbulence to decay and allowing the existence of quiescent condensations with moderate supersonic motions, similar to what we observe. A recent R-band optical polarimetry study of the Pipe by \citet{polaripipe} suggested that the mass to magnetic flux ratio toward B59 is about 1.4 (slightly supercritical), with a field strength 2 to 3 times weaker than other parts of the cloud. However, note that most of the stars available for optical polarimetry in B59 are located in regions of low column density at the edge of the core, and thus a more reliable tracer of the field towards the core could still improve these estimations.

\subsection{Fragmentation the Eastern Filament}
\label{section:discussion:subsection:grandfilament}

In the RLA09 analysis of the 2MASS extinction map, the eastern filament in B59 was considered
to be one large core because their measurement of the length and separation of the fragments detected by \texttt{CLUMPFIND-2D} did not comply with the requirements of being separated by more than a Jeans length and by more than 0.12 km/s in central velocity. Our high resolution map allows us to revisit the first requirement with a better quality dataset.

\begin{figure}
\begin{center}
\includegraphics[width=3.5in]{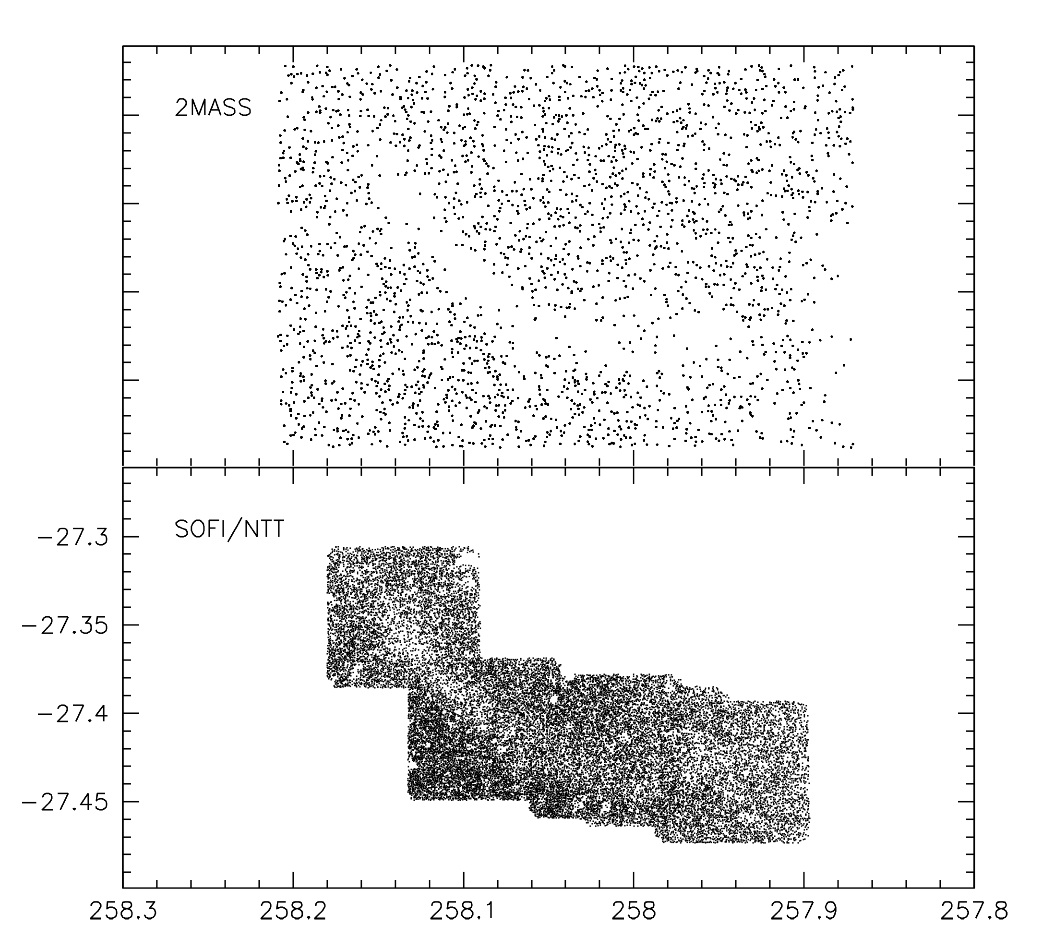}
\caption{A comparison of the density of background sources from the 2MASS (top) and the ESO (bottom) catalogs, in the region of the eastern filament. \label{fig:filpos}}
\end{center}
\end{figure}

 In Figure \ref{fig:filpos} we show the positions of sources in the area of the filament with reliable quality for application of the near-IR color excess method, from the 2MASS catalog and the ESO-NTT survey observations. The difference in the density of sources is remarkable, but not as much as the fact that the central region along the filament is almost void of sources in 2MASS. This means that LAL06 were probably not able to resolve the central part of the filament. Instead, with the ESO-NTT data we are sampling extinction all the way to the middle region. While the 2MASS map and the ESO-NTT map agree in the basic morphology, extension and basic structure of the core, the 2MASS map cannot be used to determine the presence of substructure: \citet{lal99} showed (specifically for a cylindrical cloud), that a poor sampling of the background population in a highly obscured region due to a limited photometric, depth can lead to the creation of spurious sub-structure, or at least to an unreliable determination of the size and mass of substructures. In that sense, our hybrid map resolves the filament along its axis and allows for the recognition of 
significant substructure. 

Our analysis of the filament with \texttt{CLUMPFIND-2D} reveals a total of 6 extinction peaks along the filament, and that these peaks are significant with our choice of 5 sigma contour steps. However, we must proceed with caution: again, it not trivial to claim that each one of these peaks should be considered a new core. The RLA09 prescription requires individual cores to be separated in both distance and velocity. In their analysis, RLA09 also detected substructure along the filament, but they decided to merge all peaks into a single, large core as the adjacent peaks in the map of LAL06 did not show a separation in central velocity. However, we need to consider that while the prescription of Rathborne is useful to determine the independence of two adjacent cores, it might not be ideal for a chain of related cores like in this filament. 

We suggest that the filament might be in a process of thermal fragmentation: we do not have pointed radio measurements at the peaks in the high resolution map, but we can revise the condition of the projected distance between the peaks along the filament with our improved measurements of the LJL: the small projected separation between peaks 11b and 11c is smaller than the LJL, suggesting that they might represent a single fragment, and thus we merged them in the final list of cores. Our peak to peak measured distances for the remaining features average 0.11$\pm$0.3 pc, just shorter the local Jeans length estimated for the individual peaks, which varies from 0.12 to 0.16 pc. We have to consider, however, that projection effects over such a long structure may not be small and thus our measured projected distances are lower limits.
Also, the resultant Jeans mass values, $M_J$ for the peaks along the filament average to 1.3$\pm$0.2 M$_\odot$, and the ratio $M_J/M$ has an average of 1.2$\pm$0.1, indicating that the peaks along the filament are Jeans stable. Moreover, we measured the projected width of the filament directly on the map at seven different points along the eastern filament. The average width is 0.10$\pm$0.02 pc, that compares well with the average LJL across the filament.

These measurements suggest that the eastern filament in B59 could be following a process of thermal fragmentation, and could break into several fragments. The fragmentation of isothermal, cylindrical clouds has been studied often: observations by \citet{schnelm79} suggested that in a majority of observable filaments in molecular clouds, it was possible to observe a periodicity in the separations of peaks along filaments, and that in many cases these fragments agreed spatially with the locations of young sources, suggesting that peaks observed along filaments are the precursors of independent cores. Recent observations have shown significant evidence of thermal fragmentation of filamentary clouds (e.g. \citet{wiseman98,hartmann02,lupus,paula06}).

Filamentary structures may become unstable and collapse via thermal fragmentation once they lose equilibrium with external pressure. \citet{mccrea57} found that a finite cylinder would collapse once its mass per unit length, $m=M/L$ surpasses a critical value, $m_c$. Numerical two-dimensional simulations by \citet{bastien83} allow one to make a rough estimation of the Jeans number, $J_{cyl}$, in a finite isothermal cylinder of length $L$ and diameter $D$ as $J_{cyl}\approx 3(m/m_c)f(L/D)$, where $f(L/D)$ is a geometrical factor. Following \citeauthor{bastien83} calculations, for a ratio $L/D\approx 5$ as observed in our map, the value of $f(L/D)$ from Bastien tables is close to 1.6, the $m/m_c$ ratio is about 0.3, and the critical Jeans number is $J_c$=0.6, so that $J_{cyl}/J_c =2.4$. The expected number of fragments is then roughly equal to $(J_{cyl}/J_c)^{3/2}\approx3.6$, not far from our estimate of five fragments along the eastern filament. 

\subsection{A revised list of cores in Barnard 59}
\label{section:discussion:subsection:newlist}

In table \ref{tab:newlist} we re-organize the list of map features according to the previous discussion. The table also list the locations of the cores and their peak extinction values in the unfiltered map. We confirm the seven main core identifications made by RLA09 in the map of LAL06 The analysis of our high resolution map confirms that cores 03 and 11g are single globules, while cores 04, 05, 07, 09 and 11 have significant substructure. In the case of core 11, the eastern filament, we find five fragments of significance. We suggest that peaks 11b and 11c represent one single fragment and that core 11g might not belong to the same chain. Regarding the central core we conclude that it has little observable fragmentation despite its large mass. The central core has two significant fragments, 09ab and 09c, the latter possibly being formed by a protostellar outflow; cores 09d and e are merged into a single independent core, 09de. 

\begin{deluxetable*}{lllcc}
\tablecolumns{5}
\tablewidth{0pc}
\tablecaption{Revised List of Cores in Barnard 59 \label{tab:newlist}} 
\tablehead{
\colhead{Clump ID\tablenotemark{1}} &
\colhead{RA} &
\colhead{Dec} &
\colhead{Peak ExtinctionID\tablenotemark{2}} & 
\colhead{Core Mass} \\
\cline{2-3}
\colhead{} &
\multicolumn{2}{c}{[J2000]} &
\colhead{[mag]} &
\colhead{[M$_\odot$]} \\
}
\startdata 

\multicolumn{5}{c}{Central Core Region}\\*
\cline{1-5}\\*	

09ab  & 17:11:21.8 & -27:26:10.0 & 89.0 & 18.9 \\
09c   & 17:11:27.1 & -27:23:10.0 & 29.8 &  2.8 \\

\multicolumn{5}{c}{Eastern Filament}\\*
\cline{1-5}\\*	

11a  & 17:12:32.2 & -27:21:31.8 & 20.0 & 2.6 \\
11bc & 17:12:21.9 & -27:23:49.6 & 18.3 & 1.5 \\
11d  & 17:12:09.3 & -27:25:19.8 & 15.7 & 1.7 \\
11e  & 17:11:58.8 & -27:26:22.4 & 16.2 & 1.3 \\
11f  & 17:11:50.3 & -27:25:02.5 & 16.2 & 1.0 \\

\multicolumn{5}{c}{Peripheral Cores}\\*
\cline{1-5}\\*	

03    & 17:10:31.5 & -27:25:46.6 & 22.4 & 2.4 \\
04    & 17:11:35.9 & -27:33:30.0 & 13.4 & 1.5 \\
05    & 17:12:13.7 & -27:37:19.7 & 13.1 & 1.8 \\
07    & 17:10:51.8 & -27:22:59.6 & 13.3 & 1.5 \\
09de  & 17:11:14.1 & -27:22:32.4 & 12.8 & 0.7 \\
11g   & 17:12:52.1 & -27:23:08.7 & 11.0 & 1.5 \\

\enddata
\tablenotetext{1}{Core ID after merging of features as described in text}
\tablenotetext{2}{Refers to peak extinction of core in the unfiltered map}

\end{deluxetable*}

\section{Summary}
\label{section:summary}

The main results we achieved in this study can be listed as follows:

1. Using a combination of near-infrared and mid-infrared (3 to 8 $\mu$m) observations, we obtained very deep photometry for the dense population of background stars (mostly galactic bulge giants) towards Barnard 59, the  most dense and dark region of the Pipe Nebula. The dense central core of the Barnard 59 complex is one of the very few cores in the cloud with active star formation and the only one forming a stellar cluster.

2. Our deep observations allowed us to make a detailed map of dust extinction towards the B59 region with unprecedented spatial resolution (12 to 24$\arcsec$) and dynamic range of column density ($A_V<89$ mag or $N_{H_2}<8.45\times 10^{22} \mathrm{cm}^{-2}$).

3. Our extinction map allowed us to determine, in detail, the structure of the central core as well as of 
the filaments and peripheral cores in the Barnard 59 complex. We found a total of 13 independent extinction peaks in the region, most of them interconnected by a complex structure of filaments. The largest filament is characterized by a high column density, with a peak extinction above $A_V$=22 mag. It appears to contain five independent cores evenly spaced along its axis. 

4. The total mass of the complex is approximately 41 $M_\odot$, with half of this mass (about 21 M$_\odot$) being contained in the central region (comprising cores 09ab and 09c), which is associated with the formation of the stellar cluster. 

5. The central core appears to have remarkably little structure despite the fact that its mass is significantly larger than either the Bonnor-Ebert or the Jeans mass in the region. The split of the central core into two peaks is significant, but the small projected separation of the peaks and the negligible difference in central velocities does not support the idea that these peaks represent two independent cores. Alternatively, the peaks could be related to the presence of a class 0/I source which is located between them. This source may have created a depression in what otherwise would be a large stratified core with a $1/r^2$ density distribution. The apparent separation of a fragment at the northwestern edge of the central core also appears to be significant, 
but it is at the edge of an outflow cavity and could be the result of the disruption of the core material by the stellar outflow. Our map also suggests that a second cavity is located near source No. 9. Apparently, most of the observed substructure in B59 could be the result of action by embedded YSOs, indicating that stellar feedback is playing role of significance in the evolution of the core.

7. Pointed observations of the (1,1) and (2,2) rotational transitions of NH$_3$ at 26 positions in the central core show that it is thermally dominated, with non-thermal motions being mostly subsonic. The central core has formed a small cluster of stars, but presently it is mostly quiescent and does not show
any kinematic evidence of global collapse. It is possible that the feedback from the recently formed stars may be preventing the global collapse and suppressing further fragmentation. However, the absence of a significant non-thermal component in the observed radio lines suggests that an additional source of non-thermal support is required.

8. Substructure along the axis of the eastern filament appears to be consistent with Jeans fragmentation. 

\acknowledgments

We thank Jill Rathborne for providing the reduction and model fit of the NH$_3$ data as part of a GBT survey of dense cores in the Pipe Nebula. We
thank August Muench, Kevin Covey, Doug Johnstone and Paula Teixeira for useful discussions and suggestions. This project acknowledges support from NASA Origins Program (NAG 13041), NASA Spitzer Program GO20119 and JPL contract 1279166. This work is based in part on observations made with ESO Telescopes at the La Silla and Paranal Observatories under programs 069.C-0426 and 071.C-0324. This work is based in part on observations made with the Spitzer Space Telescope, which is operated by the Jet Propulsion Laboratory, California Institute of Technology under a contract with NASA. This publication makes use of data products from the Two Micron All Sky Survey, which is a joint project of the University of Massachusetts and the Infrared Processing and Analysis Center/California Institute of Technology, funded by the National Aeronautics and Space Administration and the National Science Foundation.

{\it Facilities:} \facility{ESO (NTT) SOFI}, \facility{ESO (VLT) ISAAC}, \facility{NASA (Spitzer Space Telescope) IRAC},  \facility{GBT}.

%\multicolumn{9}{c}{Central Core Region (split)}\\*
%\cline{1-9}\\*	

%\begin{rotate}

%\end{rotate}

\end{document}